\documentclass[11pt]{article}
\usepackage{moriond,epsfig}

\def\be{\begin{equation}}
\def\ee{\end{equation}}
\def\bea{\begin{eqnarray}}
\def\eea{\end{eqnarray}}

\def\lsim{\raise0.3ex\hbox{$<$\kern-0.75em\raise-1.1ex\hbox{$\sim$}}}
\def\gsim{\raise0.3ex\hbox{$>$\kern-0.75em\raise-1.1ex\hbox{$\sim$}}}

\def\beq{\begin{equation}}
\def\eeq{\end{equation}}
\def\bea{\begin{eqnarray}}
\def\eea{\end{eqnarray}}
\def\bq{\begin{quote}}
\def\eq{\end{quote}}

\newcommand{\cc}{{c\bar{c}}}

\begin{document}
\vspace*{4cm}
\title{CHARMONIUM DISSOCIATION AND RECOMBINATION: COLD EFFECTS}

\author{
E. G. FERREIRO}

\address{
Depto. de F\'{\i}sica de Part\'{\i}culas,
Universidade de Santiago de Compostela \\
E-15782 Santiago de Compostela, Spain}

\maketitle\abstracts{
Thirty years ago, Matsui and Satz proposed the $J/\psi$ destruction as a signal of a Quark Gluon Plasma (QGP) formation, due to the Debye screening between the pair c-cbar. 
At the light of the recent experimental data, 
I review the different effects on $J/\psi$ production, from SPS to LHC energies, 
distinguishing between Cold Nuclear Matter (CNM) and QGP effects. Different possibilities and explanations for the available experimental data are discussed. Model predictions for the arriving LHC data are also presented
}

\section{Introduction}
The $J/\psi$ production constitutes one of the most puzzling fact of the present and futures experiments. Its suppression was initially proposed as a signal of the formation of a QGP. Data from NA50 collaboration at SPS energies have shown an anomalous suppression, greater than the expected one from the usual nuclear absorption. 
The data at RHIC energies from PHENIX collaboration show the same amount of suppression as the SPS data, while their energy collision is 10 times higher, and show stronger suppression at forward than at mid rapidity.

Trying to explain all these facts, 
I propose the following distinction among the effects: I will consider on one side the Cold effects, meaning by cold the fact that in those effects no thermalization is considered, 
even if the medium is dense. So these are effects without QGP.

On the other side we have the Hot effects, where thermalization is taken into account and QGP formation is included.

I will focus on the first group. Note that this group is constituted by both initial and final effects. In fact, the shadowing of the nuclear structure functions, the nuclear absorption --that is, the suppression of the $J/\psi$ due to the scattering of the pre-resonant c-cbar pair within the nucleus-- and the partonic and hadronic dissociation of the c-cbar pair with the dense medium produced in the collision --the comover interaction-- belong to this group.

\section{Shadowing} Shadowing refers to the mechanism that makes the nuclear structure functions in
nuclei different
from the superposition of those of their constituents nucleons.
The data on $d-Au$ collisions obtained by PHENIX collaboration show that there is an important suppression of the $J/\psi$ production due to this initial state effect. An exhaustive study of the nuclear modification factor versus rapidity, the number of collisions and transverse momentum has been developed in \cite{1r}. Note that the shadowing increases with $x$ and decreases with $Q^2$, so in principal its effect can be considered negligible at SPS energies. A comparison of two different types of shadowing is shown in Fig.~1 where the results from gluon shadowing \cite{2r} (CF) and EKS evolution model (EKS) are compared.  
\begin{figure}[h] 
\begin{minipage}[t]{.55\textwidth} 
\begin{center} 
\epsfig{file=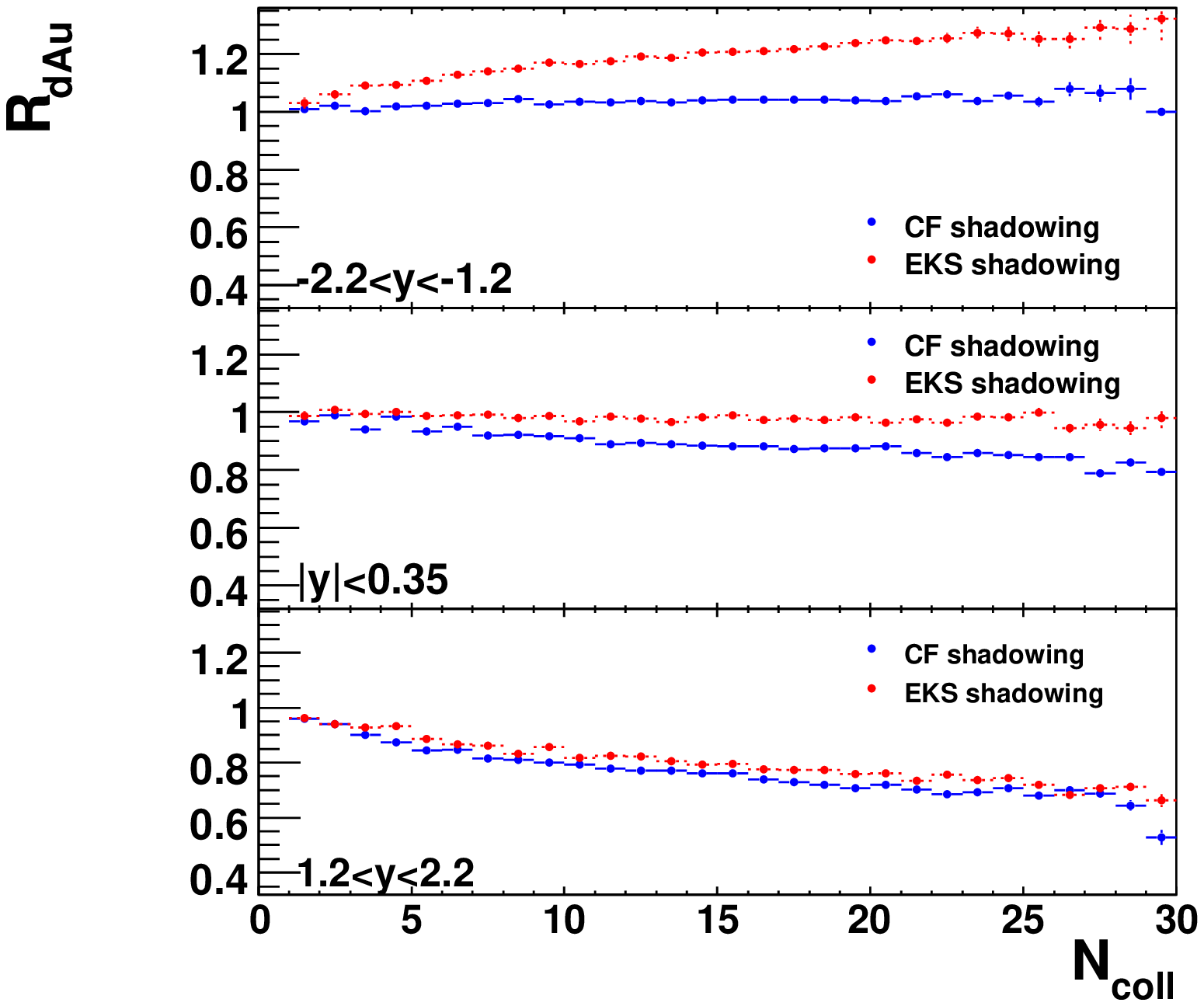, scale=0.45} 
\label{fig-tc} 
\end{center} 
\end{minipage} 
\begin{minipage}[t]{.4\textwidth} 
\begin{center} \epsfig{file=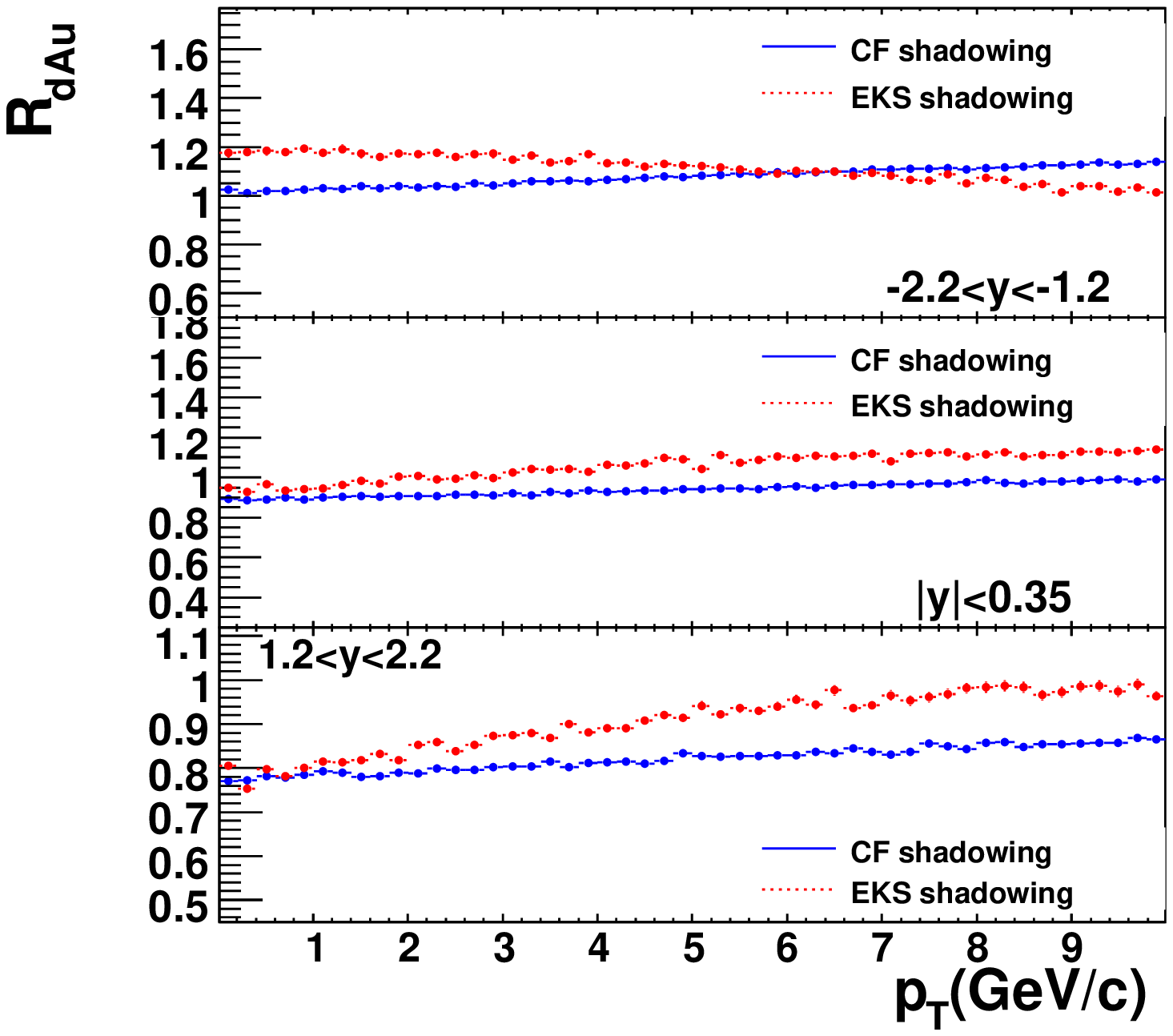, scale=0.45} 
\label{fig-tc} 
\end{center} 
\end{minipage} 
\caption{Nuclear modification factor as a function of the number of collisions (left) and as a function of the $J/\psi$ transverse momentum (right) from CF shadowing and EKS models in the absence of nuclear absorption.}
\hfill 
\end{figure}

\section{Nuclear absorption}
The primordial spectrum of particles is altered by interactions with the nuclear matter 
they traverse on the way out to the detector. This effect is known as nuclear absorption. It depends on
the coherence length, 
$$ \Delta={1 \over l_c}= m_p {M_{\cc} \over s x_1}\, ,$$
in such a way that, at low energy, the heavy system undergoes successive interactions with nucleons in
 its path and has to survive all of them leading to a strong nuclear absorption. This is the case at SPS energies, where the absorptive cross section is estimated to be $\sigma_{abs}=4.18$ mb. 
At high energies, on 
the contrary, the coherence length is large and the projectile interacts with the 
nucleus as a whole leading to a small nuclear absorption.
Then, at RHIC energy, we have taken an absorptive cross section at mid rapidity equal to zero.
Note that the coherence length shows a dependence on rapidity. Because of this, at 
forward rapidity 
a small nuclear absorption may be included.

\section{Comover interaction with recombination}
The comover interaction model was developed to explain both the
suppression of charmonium yields
and the strangeness enhancement 
in
nucleus-nucleus collisions at the SPS. 
It is based on the interaction of a particle or a parton with the medium, which is described by the
gain and loss differential equations which govern the final state interactions.

Assuming only $J/\psi$ dissociation \cite{2r}, the rate equation governing the
density of charmonium in the final state, $N_{J/\psi}$, can be written
in a simple form assuming a pure longitudinal expansion of the system and boost invariance.
For an {\it AA} collision the
density of $J/\psi$ at a given
transverse coordinate, $s$, impact parameter $b$, and rapidity is given by
\beq
\label{eq:comovrateeq}
\tau \frac{\mbox{d} N_{J/\psi}}{\mbox{d} \tau} \, \left( b,s,y \right)
\;=\; -\sigma_{co} N^{co}(b,s,y) N_{J/\psi}(b,s,y) \;,
\eeq
where $\sigma_{co}$ is the cross section of charmonium dissociation
due to interactions with the co-moving medium, with density $N^{co}$. It is found
from fits to low-energy experimental data to be $\sigma_{co} = 0.65$
mb. 

In the last years, a secondary
$J/\psi$ production due to recombination of $\cc$ pairs in order to explain the PHENIX data has been wildly discussed. 
To incorporate the effects of recombination, we have to include an
additional gain term proportional to the (squared) density of open charm
produced in the collision \cite{3r}. Then eq.~(\ref{eq:comovrateeq}) is
generalized to
\beq
\label{eq:recorateeq}
\tau \frac{\mbox{d} N_{J/\psi}}{\mbox{d} \tau} \, \left( b,s,y \right)
\;=\; -\sigma_{co} \left[ N^{co}(b,s,y) N_{J/\psi}(b,s,y) \,+\,
  N_c(b,s,y) N_{\bar{c}} (b,s,y) \right] \;,
\eeq
where we have assumed that the effective recombination cross section
is equal to the dissociation cross section.
This extension of the model
therefore does not involve additional parameters.

Equation~(\ref{eq:recorateeq})
leads to the survival probability for the $J/\psi$
\beq
\label{eq:fullsupp}
S^{co}(b,s,y) \;=\; \exp \left\{-\sigma_{co}
  \,\left[N^{co}(b,s,y)\,-\, \frac{N_c(b,s,y)
  N_{\bar{c}} (b,s,y)}{N_{J/\psi}(b,s,y)} \right] \, \ln
\left[\frac{N^{co}(b,s,y)}{N_{pp} (0)}\right] \right\} \;,
\eeq
where the
first term in the exponent of eq.~(\ref{eq:fullsupp}) is exactly the
survival probability of a $J/\psi$ interacting with comovers. 
The density of open and hidden
charm in {\it AA} collisions, $N_c,N_{\bar{c}}$ and $N_{J/\psi}$
respectively, can be computed from their densities in {\it
  pp} collisions as $N_c^{AA} (b,s) = n(b,s) S_{HQ}^{sh}(b,s)N_c^{pp}$, with similar expression for $N_{\overline{c}}^{AA}$ and $N_{J/\psi}^{AA}$. Here $n(b,s)$ corresponds to the number of collisions 
$S_{HQ}^{sh}$ is the shadowing factor
for heavy quark production.
Then
eq.~(\ref{eq:fullsupp}) becomes
\beq
S^{co}(b,s,y) \;=\; \exp \left\{-\sigma_{co} \,\left[N^{co}(b,s,y)- C(y)
    n(b,s)S_{HQ}^{sh}(b,s) \right] \, \ln
  \left[\frac{N^{co}(b,s,y)}{N_{pp} (0)}\right] \right\}
\eeq
where
\beq
\label{eq:Cratio}
C (y) \;=\; \frac{\left(\mbox{d}N^{\cc}_{pp}/\mbox{d}y
  \right)^2}{\mbox{d}
  N^{J/\psi}_{pp}/\mbox{d} y} \;=\; \frac{\left(\mbox{d}
    \sigma^{\cc}_{pp}/\mbox{d}y
  \right)^2}{\sigma_{pp} \,
  \mbox{d}\sigma^{J/\psi}_{pp}/\mbox{d} y} \;.
\eeq
We expect the effect of recombination to be stronger at
mid-rapidity than at forward ones. At $y \neq 0$ the recombination term is smaller
(relative to the
first one) since the rapidity distribution of $D$,$D^*$ is
narrower than the one of comovers.
This will produce a decrease of $R_{AA}^{J/\psi}$ with increasing $y$
which may over-compensate the increase due to a smaller density of
comovers at $y \neq 0$.

%
\begin{figure}[t!]
  \begin{center}
    \includegraphics[width=.5\linewidth]{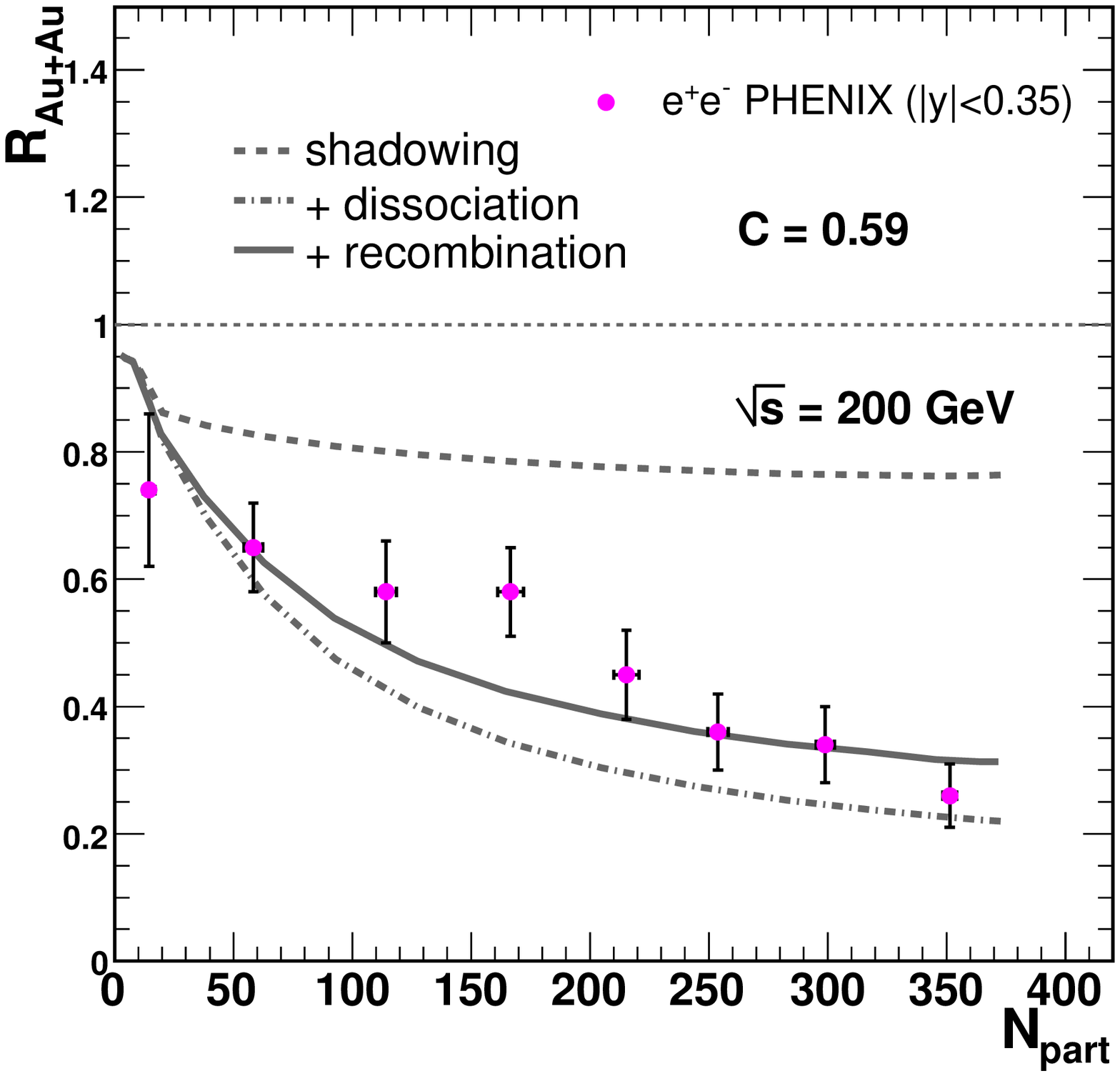}%
    \includegraphics[width=.5\linewidth]{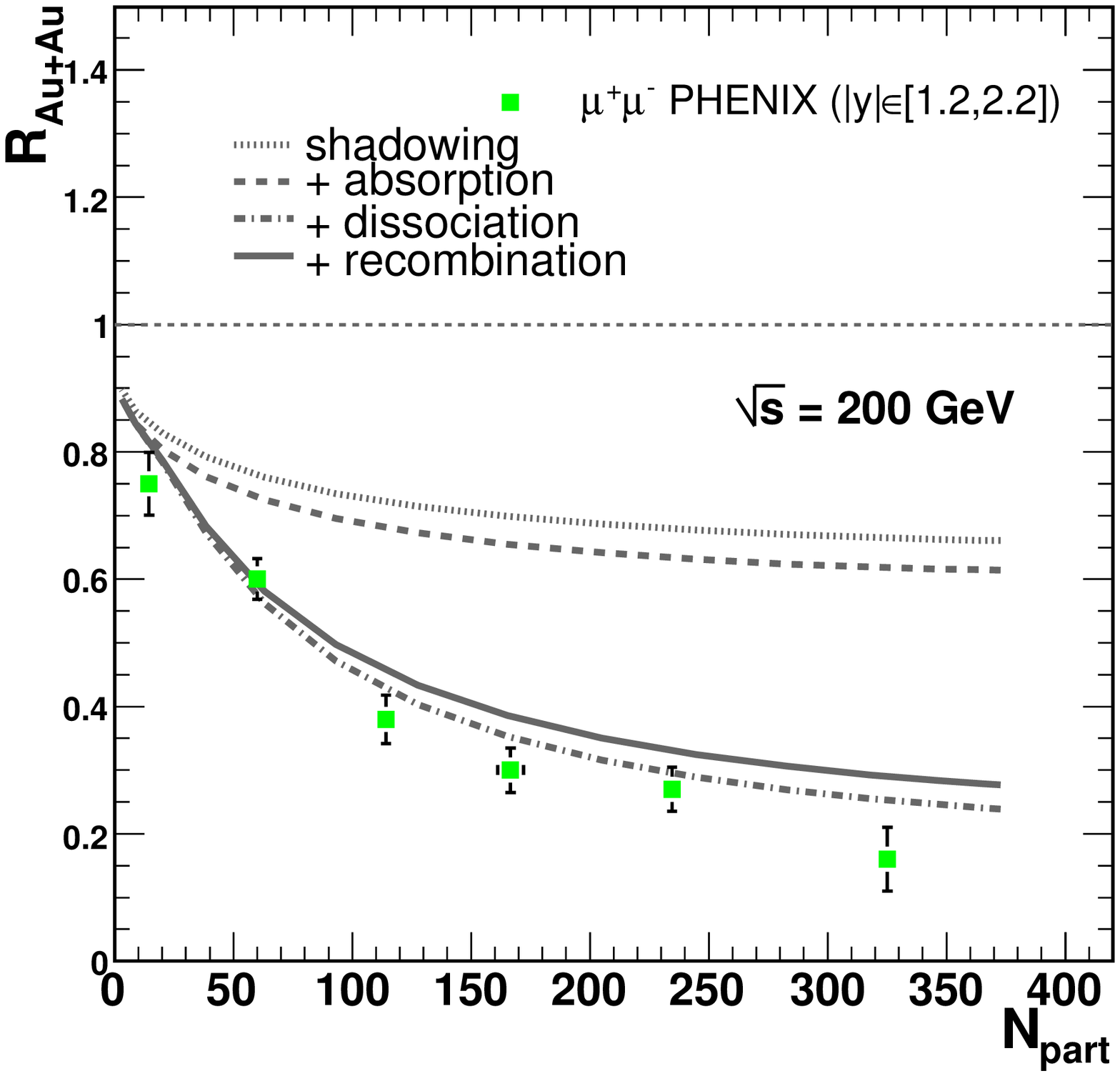}
  \end{center}
  \caption{Results for $J/\psi$ suppression in {\it AuAu} at RHIC ($\sqrt{s} = 200$ GeV) at mid-rapidity
(left),
 and at forward rapidity
    (right). Data are from {\protect\cite{4r}}.
The solid curves are the final results. The dashed-dotted ones are the results without 
recombination ($C = 0$). The dashed line is the total initial-state effect. The dotted line in the right 
figure is the result of shadowing.
}
  \label{fig:JPSIresultsRHICAuAu}
\end{figure}
In the left picture of
Fig.~2 we present the results of our
model compared to experimental data at mid-rapidity. The different
contributions to $J/\psi$ suppression are shown. 
Note that at mid-rapidities the initial-state effect is just the
shadowing. As discussed above, nuclear absorption 
is present at forward rapidities but is
negligibly small at mid-rapidities.
In the r.h.s. of Fig.~2, our results at forward rapidity are presented. Note that, contrary to the results in \cite{2r}
with no recombination, the $J/\psi$ suppression at forward rapidity is
somewhat larger that the one at mid-rapidities, in agreement with
experimental data. This is due both to the recombination term and to
the initial-state effects. The latter are stronger for forward
rapidities.

In Fig.~3 we have calculated the $J/\psi$
suppression at LHC
for several values of C, including the case of absence of
recombination effects ($C=0$).
We
consider that realistic values of $C$ at LHC are of the range 2 to 3.
Although the density of charm grows substantially from RHIC to LHC,
the combined effect of initial-state shadowing and comovers
dissociation appears to overcome the effect of parton recombination.
This is in sharp contrast with some of the findings in the framework of QGP and recombination,
where
a strong enhancement of the $J/\psi$ yield with increasing centrality
is predicted.
\begin{figure}[t!]
  \begin{center}
    \includegraphics[scale=0.4]{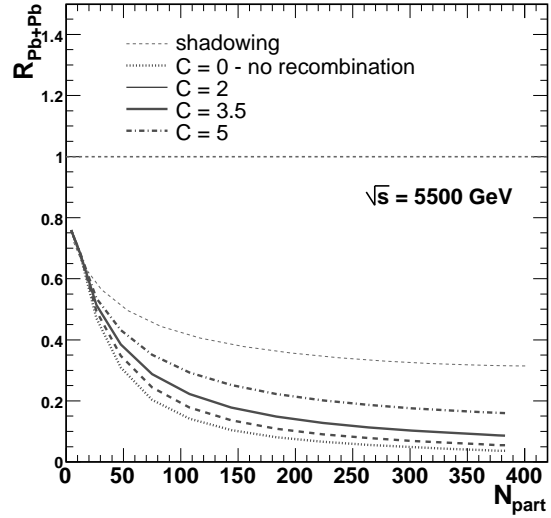}
  \end{center}
  \caption{Results for $J/\psi$ suppression in {\it PbPb} at LHC ($\sqrt{s} = 5.5$
    TeV) at mid-rapidities for different values
    of the parameter $C$. The upper line is the suppression due to initial-state effects (shadowing).}
  \label{fig:JPSIresultsLHC}
\end{figure}

\end{document}